\documentstyle[aps,preprint]{revtex}

\begin{document}

\draft

\preprint{\vbox{ \hbox{SOGANG-HEP 240/98} \hbox{SNUTP 98-072}}}

\title{Symplectic structure free Chern-Simons Theory}

\author{Won Tae Kim, Yong-Wan Kim and Young-Jai Park}

\address{Department of Physics 
and Basic Science Research Institute,\\
Sogang University, C.P.O. Box 1142, Seoul 100-611, Korea}

\date{July 23}

\maketitle

\begin{abstract}
The second class constraints algebra of the abelian Chern-Simons theory 
is rigorously studied in terms of the
Hamiltonian embedding in order to obtain the first
class constraint system. The symplectic structure of fields 
due to the second
class constraints disappears in the resulting system. 
Then we obtain a new type of 
Chern-Simons action which has an infinite set of the irreducible first class 
constraints and exhibits new 
extended local gauge symmetries implemented by
these first class constraints.
\end{abstract}

\pacs{PACS number(s): 11.10.Ef, 11.10.Kk.}


\newpage
Chern--Simons (CS) theories \cite{djt} have been enormously
studied in the various arena.   
One of the intriguing problems of CS theories is 
that CS Lagrangian from the point of view of 
constrained system  
gives unusual second class constraints even though it is
invariant up to a total divergence under the local  
gauge transformation. 
This peculiar property of CS theories is essentially due to 
the symplectic structure \cite{fj} which is a key ingredient of
CS systems. Meanwhile second class constraint system 
has been generically regarded as
a gauge fixed version of gauge invariant system \cite{fs}, which
has been studied in the context of anomalous gauge theory \cite{jr}.
Therefore CS theories may have some kinds of unknown 
local symmetry if second class constraints are converted into
first class ones. This means that the intrinsic symplectic 
structure of CS theories can be interpreted as a gauge fixed form of
gauge invariant theory which is symplectic free. Then what is
the additional local symmetry in connection with 
the symplectic structure? 
    
It is in general difficult to convert the
second class algebra even for the abelian pure
CS theory into first class constraint system
by using the St\"uckelberg mechanism \cite{ST}
since the origin of the second class constraints algebra is unusual
compared to the conventional anomalous theory.
Therefore Batalin, Fradkin, and Tyutin (BFT)  
Hamiltonian embedding of a model \cite{bft} is very useful,
which converts systematically second class constrained system into
first class one. 
According to the usual treatment \cite{ban,kp,brr,kpkk,bb} 
of the BFT formalism, 
one simply identifies auxiliary fields with a pair of conjugate fields.  
This procedure in the CS theory however gives an 
undesirable final expression 
\cite{ban,kpkk} in that the original action has not been reproduced 
when we choose the unitary gauge \cite{ban,kp}, and the additional
action so called Wess--Zumino (WZ) action,
which is needed to make gauge invariant system, is model--dependent. 
Furthermore the assumed 
brackets of the auxiliary fields are not Poisson brackets but Dirac ones.   

In this paper by introducing infinite auxiliary
fields, we find a new type of WZ action for the abelian pure CS theory 
so that the total system has fully 
first class constraints which have not been successful so far. 
This total action is naturally reduced to the
original CS action if one chooses the unitary gauge conditions.
Then we obtain the new symmetries  corresponding to
the first class constraints related to the symplectic structure as well as
the well--known local U(1) gauge symmetry.

Let us start with the abelian pure CS Lagrangian
\begin{equation}
\label{1}
{\cal L}_0=\frac{\kappa}{2}\epsilon_{\mu\nu\rho}A^\mu\partial^\nu A^\rho.
\end{equation}
The canonical momenta are given by $\pi_0=0$ and $\pi_i=\frac{\kappa}{2}
\epsilon_{ij}A^j$. Then we have three primary constraints \cite{dir},
$\Omega_0 = \pi_0 \approx 0$,
$\Omega_i = \pi_i-\frac{\kappa}{2}\epsilon_{ij}A^j \approx 0~ (i,j=1,2)$,
and a secondary constraint as
\begin{equation}
\label{4}
\Omega_3 = \kappa \epsilon_{ij}\partial^iA^j \approx 0,
\end{equation}
which is obtained from the stability condition of time evolution of the 
$\Omega_0$ with the primary Hamiltonian
$H_p = H_c + \int d^2x (v^0\Omega_0 + v^i\Omega_i)$,
where the canonical Hamiltonian is 
$H_c=\int d^2x {\cal H}_c=\int d^2x \ \kappa A^0\epsilon^{ij}\partial_iA^j$.
No more additional constraints are generated from the consistency conditions
of the other constraints $\Omega_i$ and $\Omega_3$ by fixing the Lagrange 
multipliers as $v^i=\kappa\partial^iA^0$. 

To obtain the maximally irreducible first class constraints 
\cite{ban},
we redefine the above primary and secondary constraints as
$\omega_0 \equiv \Omega_0$,
$\omega_i \equiv \Omega_i$, and
\begin{eqnarray}
\label{8}
\omega_3 \equiv \Omega_3 + \partial^i\Omega_i 
             = \partial^i\pi_i + \frac{\kappa}{2}\epsilon_{ij}\partial^iA^j
               \approx 0.
\end{eqnarray}
Eliminating the Lagrange multipliers $v^i$ yields 
the total Hamiltonian density \cite{ht} corresponding to the CS Lagrangian 
\begin{equation}
\label{9}
{\cal H}_T = v^0\omega_0-(A^0-v^3)\omega_3,
\end{equation}
where the Lagrange multipliers $v^0$ and $v^3$ remain  
undetermined. The total Hamiltonian now naturally generates 
the Gauss' constraint $\omega_3$ from the time evolution of the 
constraint $\omega_0$ as
\begin{eqnarray}
\label{10}
\dot\omega_0 = \omega_3,~~
\dot\omega_\alpha = 0~~~~(\alpha=1,2,3),
\end{eqnarray}
where overdot represents the time evolution.
We therefore have two first class constraints $\omega_0$ and $\omega_3$, and
two second class constraints $\omega_i$ which satisfy the constraint algebra
\begin{eqnarray}
\label{11}
\Delta_{ij}(x,y) \equiv 
\{\omega_i(x), \omega_j(y)\} = -\kappa\epsilon_{ij}\delta(x-y).
\end{eqnarray}
Upon elimination of the momenta $\pi_i$ via the method of Dirac \cite{dir},
we could easily obtain the well--known Dirac brackets for the gauge fields 
$A^i$ as
$\{A^i(x), A^j(y)\}=\epsilon^{ij}\delta(x-y)/\kappa$.

Compared to this kind of the phase space reduction, 
one can embed a second class structure into first class one 
by introducing auxiliary fields via 
{\it a la} BFT Hamiltonian embedding \cite{bft}.
In order to make the analysis explicitly, let us first rewrite 
the CS Lagrangian replacing $A^\mu$ with $A^{(0)\mu}$ as 
\begin{equation}
\label{13}
{\cal L}_0 \equiv {\cal L}^{(0)} =
             \frac{\kappa}{2}\epsilon_{\mu\nu\rho}A^{(0)\mu}
                     \partial^\nu A^{(0)\rho}.
\end{equation}
We now introduce auxiliary fields $A^{(1)i}$ to make the second class
constraints $\omega_i$ into first class ones satisfying
$\{A^{(1)i}(x),A^{(1)j}(y)\}=\vartheta^{ij}(x,y)$.
Making use of the auxiliary fields $A^{(1)i}$, we could write the effective
first class constraints as 
$\tilde{\omega}_i (\pi^{(0)}_\mu,A^{(0)\mu};A^{(1)i}) 
             = \omega_i + \sum_{n}\varpi_i^{(n)}$
satisfying the boundary condition 
$\tilde{\omega}_i(\pi^{(0)}_\mu,A^{(0)\mu};0)=\omega_i$ 
as well as requiring the strong involution, {\it i.e.,} 
$\{\tilde{\omega}_i,\tilde{\omega}_j\}=0$.
Here $\varpi_i^{(n)}$ is assumed to be proportional to $(A^{(1)i})^n$.
In particular, the first order correction in these infinite series is given by
\begin{equation}
\label{16}
\varpi^{(1)}_i = \int \ d^2y \ X_{ij}(x,y)A^{(1)j},
\end{equation}
and the requirement of the strong involution gives the condition of
\begin{equation}
\label{17}
\Delta_{ij}+
\int \ d^2ud^2v \ X_{ik}(x,u)\vartheta^{k\ell}(u,v)X_{j\ell}(v,y)=0.
\end{equation}
We take the simple solution of $\vartheta^{ij}$ and $X_{ij}$ as
\begin{eqnarray}
\label{18}
\vartheta^{ij}(x,y) &=& \epsilon^{ij}\delta(x-y), \\
\label{19}
X_{ij}(x,y) &=& -\epsilon_{ij}\delta(x-y)/\sqrt\kappa.
\end{eqnarray}
There are some arbitrariness in choosing 
$\vartheta^{ij}$ and $X_{ij}$ from (\ref{17}) as shown in the literature
\cite{ban,kp,brr},
which are related to a canonical transformation each other.
For the case (\ref{18}), Eq. (\ref{17}) is simply reduced to
$|detX_{ij}(x,y)|=1/\kappa$ and the convenient solution is
chosen just like Eq. (\ref{19}).

By using the $\vartheta^{ij}$ and solution of $X_{ij}$ 
in Eqs. (\ref{18}) and (\ref{19}),
we obtain the strongly involutive first class constraints 
which are proportional only to the first order of the auxiliary fields as
\begin{equation}
\label{20}
\tilde{\omega}^{(0)}_i = \pi^{(0)}_i - \frac{\kappa}{2}\epsilon_{ij}A^{(0)j}
                   - \sqrt{\kappa}\epsilon_{ij}A^{(1)j}=0,
\end{equation}
and the canonical Hamiltonian density
\begin{equation}
\label{21}
\tilde{\cal H}_c =  \kappa A^{(0)0}\epsilon^{ij}\partial_i
                \left(A^{(0)j} + \frac{1}{\sqrt{\kappa}}A^{(1)j}\right),
\end{equation}
satisfying $\{\tilde{\omega}_i, \tilde{H}_c\}=0$ and
$\{\omega_0,\tilde{H}_c\}=\{\omega_3,\tilde{H}_c\}=0$.
The corresponding Lagrangian of Eq. (\ref{21}) with two auxiliary
fields $A^{(1)i}$ is obtained through the usual path integral as follows
\begin{eqnarray}
\label{22}
{\cal L}^{(1)}&=&-\frac{\kappa}{2}\epsilon_{ij} A^{(0)i}\dot{A}^{(0)j}
                 + \kappa A^{(0)0}\epsilon_{ij}\partial^iA^{(0)j} \nonumber\\
              && - \frac{1}{2}\epsilon_{ij} A^{(1)i}\dot{A}^{(1)j}
                  +\sqrt{\kappa} A^{(0)0} \epsilon_{ij} \partial^i A^{(1)j} 
                  -\sqrt{\kappa}\epsilon_{ij}A^{(1)i}\dot{A}^{(0)j}.
\end{eqnarray}
Note that in the usual BFT Hamiltonian embedding of the model 
we would have
identified two auxiliary fields with a pair of conjugate fields as
coordinate and momenta \cite{ban,kp,brr,kpkk,bb}. 
However, it is problematic in our model in that
there is no priority to choose them here as the conjugate fields.

To make this problem explicitly, let us now study whether or not
the Lagrangian 
(\ref{22}) gives first class constraint system at the 
Poisson bracket level which is an essential spirit of the BFT formalism.
The canonical momenta from (\ref{22}) are $\pi^{(0)}_0=0$, 
$\pi^{(0)}_i=\frac{\kappa}{2}\epsilon_{ij}A^{(0)j}+\sqrt{\kappa}
\epsilon_{ij}A^{(1)j}$, and
$\pi^{(1)}_i=\frac{1}{2}\epsilon_{ij}A^{(1)j}$. 
From the time stability conditions of these primary constraints, 
we can get one more secondary constraint and after redefining
the constraints we can easily obtain the maximally irreducible first class 
constraints as
$\omega_0=\pi^{(0)}_0 \approx 0$,
$\omega_3 = \partial^i\pi^{(0)}_i 
              + \frac{\kappa}{2}\epsilon_{ij}\partial^iA^{(0)j} \approx 0$,
and
\begin{eqnarray}
\label{24}
\tilde\omega^{(1)}_i=\pi^{(0)}_i
                -\frac{\kappa}{2}\epsilon_{ij}A^{(0)j}
                -\sqrt{\kappa}(\pi^{(1)}_i+\frac{1}{2}\epsilon_{ij}A^{(1)j}) \approx 0,
\end{eqnarray}
as well as the second class constraints 
\begin{eqnarray}
\label{25}
\omega^{(1)}_i = \pi^{(1)}_i-\frac{1}{2}\epsilon_{ij}A^{(1)j} \approx 0.
\end{eqnarray}
Therefore there remains still second class constraint system even
after the first order of correction. On the other hand,
we can calculate the (preliminary) Dirac brackets as
\begin{equation}
\label{26}
\{A^{(1)i}, A^{(1)j}\}_D = \epsilon^{ij}\delta(x-y),
\end{equation}
which are nothing but the relation introduced in Eq. (\ref{18}) 
to make the second class constraints $\omega_i$ into first class ones 
$\tilde\omega^{(0)}_i$ in the BFT formalism,
{\it i.e.}, the first class constraints $\tilde\omega^{(1)}_i$ reduce to the $\tilde\omega^{(0)}_i$.  
As a result we observe that the auxiliary fields $A^{(1)i}$ introduced to
make the second class constraints into first ones do not provide
the Poisson bracket structure but Dirac one.
The similar feature has appeared in the chiral boson theory \cite{ww,mwy}
and recently string and D-branes theory \cite{dpa}.
This makes the BFT Hamiltonian embedding of the CS theory not stopping any 
finite number of steps. 
Therefore these steps should infinitely repeated, and thus 
we can construct an action which now has fully first class constraints
by introducing infinite auxiliary fields denoted by $A^{(n)i}$.
In this respect, all the previous results \cite{ban,kp,kpkk}
of the BFT formalism applied to 
the CS cases are incomplete.
Hence the final action can be written in the form of
\begin{eqnarray}
\label{27}
{\cal L} &=& -\frac{\kappa}{2}\epsilon_{ij} A^{(0)i}\dot{A}^{(0)j}
             + \kappa A^{(0)0}\epsilon_{ij}\partial^iA^{(0)j} \nonumber\\
          && - \frac{1}{2}\epsilon_{ij}\sum^{\infty}_{n=1}
              A^{(n)i}\dot{A}^{(n)j}            
             + \sqrt\kappa A^{(0)} \epsilon_{ij}\sum^{\infty}_{n=1}
              \partial^i A^{(n)j} \nonumber\\
          && -\sqrt\kappa\epsilon_{ij}\sum^{\infty}_{n=1} A^{(n)i}\dot A^{(0)j}
             -\epsilon_{ij}\sum^{\infty}_{n=1}\sum^{\infty}_{m=n+1}
               A^{(m)i}\dot A^{(n)j}.
\end{eqnarray}

To examine whether or not the action (\ref{27}) gives really first
class constraint system, we should check the constraint algebra 
by using the Poisson brackets.  
The canonical momenta from (\ref{27}) are given by
\begin{eqnarray}
\label{28}
\pi^{(0)}_0&=&0, \nonumber \\
\pi^{(0)}_i&=&\frac{\kappa}{2}\epsilon_{ij}A^{(0)j}
             +\sqrt{\kappa}\epsilon_{ij}\sum^{\infty}_{n=1}A^{(n)j}, 
\nonumber\\
\pi^{(n)}_i&=&\frac{1}{2}\epsilon_{ij}A^{(n)j}
             +\epsilon_{ij}\sum^{\infty}_{m=n+1}A^{(m)j},
\end{eqnarray}
where $n=1,2,\cdot\cdot\cdot,\infty$.
We thus have primary constraints as
\begin{eqnarray}
\label{29}
\Omega_0&=&\pi^{(0)}\approx 0, \nonumber\\
\Omega^{(0)}_i&=&\pi^{(0)}_i-\frac{\kappa}{2}\epsilon_{ij}A^{(0)j}
             -\sqrt{\kappa}\epsilon_{ij}\sum^{\infty}_{n=1}A^{(n)j} \approx 0, 
\nonumber \\
\Omega^{(n)}_i&=&\pi^{(n)}_i-\frac{1}{2}\epsilon_{ij}A^{(n)j}
             -\epsilon_{ij}\sum^{\infty}_{m=n+1}A^{(m)j} \approx 0,
\end{eqnarray}
whose time stability conditions give one further constraint as
\begin{equation}
\label{30}
\Omega_3=\kappa\epsilon_{ij}\partial^i
        \left(A^{(0)j}+\frac{1}{\sqrt{\kappa}}
                \sum^{\infty}_{n=1}A^{(n)j} \right) \approx 0,
\end{equation}
with the primary Hamiltonian density
\begin{equation}
\label{31}
{\cal H}_p={\cal H}_c+v^0\Omega_0+\sum_{n,i}v^{(n)i}\Omega^{(n)}_i,
\end{equation}
where the canonical Hamiltonian density is given by
${\cal H}_c=\kappa A^{(0)0}\epsilon^{ij}
\partial_i(A^{(0)j}+\frac{1}{\sqrt{\kappa}}\sum_{n=1}A^{(n)j})$.
The constraint (\ref{30}) is obtained only from the evolution of the
$\Omega_0$ and the other primary constraints do not generate any further 
constraints.
The maximally irreducible first class constraints are now obtained from 
redefining the constraints (\ref{29}) and (\ref{30}) as
$\omega_0=\pi^{(0)}_0 \approx 0$,
$\omega_3=\partial^i\pi^{(0)}_i+\frac{\kappa}{2}\epsilon_{ij}
           \partial^iA^{(0)j} \approx 0$, and
\begin{eqnarray}
\label{34}
&&\tilde{\omega}^{(1)}_i=\pi^{(0)}_i-\frac{\kappa}{2}\epsilon_{ij}A^{(0)j}
          -\sqrt{\kappa}(\pi^{(1)}_i+\frac{1}{2}\epsilon_{ij}A^{(1)j})
           \approx 0, \nonumber \\
&&\tilde{\omega}^{(n+1)}_i=\pi^{(n)}_i-\frac{1}{2}\epsilon_{ij}A^{(n)j}
          -(\pi^{(n+1)}_i+\frac{1}{2}\epsilon_{ij}A^{(n+1)j})
           \approx 0,  
\end{eqnarray}
where $n=1,2,\cdot\cdot\cdot\infty$, 
and the total Hamiltonian density has the form of
\begin{equation}
\label{36}
\tilde{\cal H}_T=\lambda^0\omega_0-(A^{(0)0}-\lambda^3)\omega_3
           +\sum^{\infty}_{n=1}\lambda^{(n)i}\tilde{\omega}^{(n)}_i.
\end{equation}
The above constraints are all involutive,
\begin{eqnarray}
\label{37}
\{\omega_0, \tilde{H}_T\}&=&\omega_3,~~
\{\omega_3, \tilde{H}_T\}=0,~~
\{\tilde{\omega}^{(n)}_i, \tilde{H}_T\}=0.
\end{eqnarray}
So the new CS theory with the infinite auxiliary fields now completely 
forms the first class constrained system 
and strongly vanishing Poisson brackets between the constraints, $\omega_0$,
$\omega_3$, and (\ref{34}). 

It seems to be appropriate to comment on the constraints, $\omega_0$,
$\omega_3$, and (\ref{34}). 
The constraint $\omega_3$ is the usual Gauss' constraint
related to the time independent gauge transformation and it is not 
modified through the BFT procedure, which reflects the
maintenance of the well--known original U(1) gauge symmetry.
On the other hand, the infinite number of  
the first class constraints (\ref{34}) are related to a kind of
unknown local symmetries and  the symplectic structure of 
the original fields is regarded as a gauge fixed structure of
the modified CS theory (\ref{27}).

Now we are ready to discuss  new local 
symmetries of our first class action.
The first order form of the action is described as
\begin{equation}
\label{38}
S=\int \ d^3x \left( \pi^{(0)}_0\dot{A}^{(0)0}
             +\sum^{\infty}_{n=0}\pi^{(n)}_iA^{(n)i}
             -\tilde{\cal H}_T \right).
\end{equation}
This action is invariant under the following gauge transformations
\begin{eqnarray}
\label{39}
\delta A^{(0)0}&=&\epsilon^0,\nonumber\\
\delta A^{(0)i}&=&-\partial^i\epsilon^3
                +\frac{1}{\sqrt\kappa}\epsilon^{(1)i}, \nonumber\\
\delta A^{(n)i}&=&-\epsilon^{(n)i}+\epsilon^{(n+1)i}, \nonumber \\
\delta\pi^{(0)}&=&0,\nonumber\\
\delta\pi^{(0)}_i&=&-\frac{\kappa}{2}\epsilon_{ij}\partial^j\epsilon^3
           -\frac{\sqrt\kappa}{2}\epsilon_{ij}\epsilon^{(1)j}, \nonumber\\
\delta\pi^{(n)}_i&=&-\frac{1}{2}\epsilon_{ij}
            \left( \epsilon^{(n)j}+\epsilon^{(n+1)j} \right), \nonumber\\
\delta\lambda^0 &=& \dot\epsilon^0,\nonumber\\
\delta\lambda^{(1)i}&=& \frac{1}{\sqrt\kappa}\dot\epsilon^{(1)i},~~
\delta\lambda^{(n+1)i}= \dot\epsilon^{(n+1)i},\nonumber\\
\delta\lambda^3&=&\epsilon^0+\dot\epsilon^3~~~~~~(n=1,2,\cdot\cdot\cdot),
\end{eqnarray}
which are generated from the definition of the gauge transformation
generators as
\begin{equation}
\label{40}
G=\int \ d^2x \left(\epsilon^0\omega_0+\epsilon^3\omega_3
             + \frac{1}{\sqrt{\kappa}}\epsilon^{(1)i}\tilde\omega^{(1)}_i
             +\sum^{\infty}_{n=2}\epsilon^{(n)i}\tilde{\omega}^{(n)}_i \right),
\end{equation}
with the infinitesimal gauge parameters 
$\epsilon^0$, $\epsilon^3$, and 
$\epsilon^{(n)i}~(n=1,2,\cdot\cdot\cdot\infty)$ 
where we have inserted $1/\sqrt{\kappa}$ in front of 
the parameters $\epsilon^{(1)i}$ for convenience. 
The equations of motion of the Lagrange multipliers $\lambda^{(n)i}$ give
the constraints $\tilde{\omega}^{(n)}_i$, while the $\lambda^{(n)i}$ themselves
can be gauged away.
Upon the gauge condition of $\lambda^3=0$ and thus $\delta\lambda^3=0$
\cite{ht},
the action (\ref{38}) reduces to
\begin{equation}
\label{41}
S=\int \ d^3x \left( \pi^{(0)}_0\dot{A}^{(0)0}
             +\sum^{\infty}_{n=0}\pi^{(n)}_iA^{(n)i}
            -\lambda^0\omega_0+A^{(0)0}\omega_3
             -\sum^{\infty}_{n=1}\lambda^{(n)i}\tilde{\omega}^{(n)}_i \right)
\end{equation}
by identifying $\epsilon^0=-\dot\epsilon^3$.
Note that the partially gauge fixed action (\ref{41}) is invariant 
under the residual gauge transformations.
To be exhausted all the additional gauge degrees of freedom, we choose gauge conditions
as $\chi^{(n)i}=\pi^{(n)}_i+\frac{1}{2}\epsilon_{ij}A^{(n)j}\approx 0,~(n=1,
2,\cdot\cdot\cdot)$ with Eq. (\ref{34}) and $\lambda^{(n)i}=0$
together, similarly to the case of chiral boson \cite{mwy}. 
We can therefore recover the original pure CS Lagrangian (\ref{1})
remaining only the usual U(1) gauge symmetry.

On the other hand, if one eliminates $\pi^{(0)}_0$, $\pi^{(0)}_i$,
$\pi^{(n)}_i$, $\lambda^0$, and $\lambda^{(n)i}~ (n=1,2,\cdot\cdot\cdot)$ 
from the action (\ref{41}) by means of their own equations of motion, 
we could get once again the desired action (\ref{27}),
and compactly rewrite it as follows
\begin{eqnarray}
\label{42}
{\cal L}   &=&-\frac{\kappa}{2}\epsilon_{ij}
    \left( A^{(0)i}+\frac{1}{\sqrt{\kappa}}\sum^{\infty}_{n=1}A^{(n)i} \right)
    \left( \dot{A}^{(0)j}
          +\frac{1}{\sqrt{\kappa}}\sum^{\infty}_{n=1}\dot{A}^{(n)j} \right)
\nonumber \\
&&    +\kappa A^{(0)0}\epsilon_{ij}\partial^i
\left( A^{(0)j}+\frac{1}{\sqrt{\kappa}}\sum^{\infty}_{n=1}A^{(n)j} \right). 
\end{eqnarray}
Then one can easily check that 
this action is invariant under the following local gauge 
transformations of
\begin{eqnarray}
\label{43}
\delta A^{(0)0}&=&\partial^0\Lambda,\nonumber\\
\delta A^{(0)i}&=&\partial^i\Lambda
            +\frac{1}{\sqrt\kappa}\epsilon^{(1)i}, \nonumber\\
\delta A^{(n)i}&=&-\epsilon^{(n)i}+\epsilon^{(n+1)i}~~(n=1,2,\cdot\cdot\cdot),
\end{eqnarray}
where we simply defined $\epsilon^3=-\Lambda$. 
The transformation rules imply that
the usual U(1) gauge transformation with the gauge parameter $\Lambda$
and a new type of local symmetries with $\epsilon^{(n)i}$. 

In conclusions, 
we have found a new type of the WZ action for the abelian pure CS theory.
To make two initial second class constraints which are originated from the 
symplectic structure into the first class system, 
we have introduced infinite number of the auxiliary fields 
via the BFT formalism. 
It is remarkable that not only the original U(1) gauge symmetry 
is still preserved but also there exist the additional novel symmetries.
Further, the derived WZ action is eventually independent of field theoretic
models which involve the CS term.
If nonabelian CS theory is considered in this way, some 
various applications may be possible, for example, the
boundary conformal field theory of the CS theory and recent
study of black hole physics.  


\acknowledgments

We would like to thank S. J. Yoon for helpful discussions.
The present study was supported by
the Basic Science Research Institute Program,
Ministry of Education, Project No. BSRI--98--2414.
WTK is in part supported by Korea Science and Engineering Foundation through
the Center for Theoretical Physics in Seoul National University (1998).



\begin{references}
\bibitem{djt} S. Deser, R. Jackiw and s. Templeton, Ann. of Phys. 
                  {\bf 140}, 372 (1982); E. Witten, Commun. Math. Phys. 
                  {\bf 121}, 351 (1989); G. Moore and N. Seiberg, Phys. Lett.
                  {\bf B220}, 422 (1989); S. Elitzur, G. Moore, A. Schwimmer
                  and N. Seiberg, Nucl. Phys. {\bf B326}, 108 (1989);
                  S. Carlip, {\it Lectures on (2+1) Dimensional Gravity},
                  gr--qc/9503024;
                  M. Ba\~nados, Phys. Rev. {\bf D52}, 5816 (1996).
\bibitem{fj} R. Floreanini and R. Jackiw, Phys. Rev. Lett. {\bf 60},
                  1692 (1988). 
\bibitem{fs} L. D. Faddeev and S. L. Shatashivili, Phys. Lett. {\bf B167}, 
                 225 (1986).
\bibitem{jr} R. Jackiw and R. Rajaraman, Phys. Rev. Lett. {\bf
                  54}, 1219 (1985).
\bibitem{ST} E. G. St\"uckelberg, Helv. Phys. Acta. {\bf 30}, 209 (1957).
\bibitem{ban} R. Banerjee, Phys. Rev. {\bf D48}, R5467 (1993).
\bibitem{kp} W. T. Kim and Y.--J. Park, 
                 Phys. Lett. {\bf B336}, 376 (1994). 
\bibitem{bft} I. A. Batalin and E. S. Fradkin, Phys. Lett.
            {\bf B180}, 157 (1986); Nucl. Phys. {\bf B279}, 514 (1987);
             I. A. Batalin and I. V. Tyutin, Int. J. Mod. Phys.
             {\bf A6}, 3255 (1991).
\bibitem{brr} R. Banerjee, H. J. Rothe and K. D. Rothe, Phys. Rev. 
           {\bf D52}, 3750 (1995); {\it ibid}., {\bf D55}, 6339 (1997).
\bibitem{kpkk} Y.--W. Kim, Y.--J. Park, K. Y. Kim and Y. Kim, Phys. Rev. 
                   {\bf D51}, 2943 (1995).
\bibitem{bb}  R. Banerjee and J. Barcelos--Neto, Nucl. Phys. {\bf B499}, 
              453 (1997); Yong-Wan Kim and K. D. Rothe, Nucl. Phys. {\bf B511},
              510 (1998); W. Oliveira and J. A. Neto, 
              to appear in Nucl. Phys., hep--th/9803258.
\bibitem{dir} P. A. M. Dirac, {\it Lectures on quantum mechanics}
         (Belfer graduate School, Yeshiba University Press, New York, 1964).
\bibitem{ht} M. Henneaux and C. Teitelboim, 
               {\it Quantization of Gauge systems}
               (Princeton University Press, Princeton, New Jersey, 1992).
\bibitem{ww} C. Wotzasek, Phys. Rev. Lett. {\bf 66}, 129 (1991).
\bibitem{mwy} B. McClain, Y.--S. Wu and F. Yu, Nucl. Phys. {\bf B343},
                 689 (1990); F. Devecchi and M. Henneaux, 
                Phys. Rev. {\bf D54}, 1606 (1996). 
\bibitem{dpa} P. Pasti, D. Sorokin and M. Tonin, Phys. Rev. {\bf D52}, 4277
               (1995); M. Aganagic, J. Park, C. Popescu and J. Schwarz,
               Nucl. Phys. {\bf B496}, 191 (1997).
\end{references}
\end{document}